\documentclass[10pt,conference,twocolumn]{IEEEtran}
\usepackage{cite}
\usepackage{times}
\usepackage{algpseudocode}
\usepackage{mathtools}
\usepackage{amsmath}
\usepackage{booktabs}
\usepackage{graphicx}
\usepackage[figuresright]{rotating}
\usepackage{pict2e}
\usepackage{algorithm}
\usepackage{amssymb,amsmath,amsthm}
\usepackage{amsmath}
\usepackage{setspace}
\usepackage{comment}
\usepackage{balance}
\usepackage{rotating}
\usepackage{multirow}
\usepackage{wrapfig}
\usepackage{mathtools}
\usepackage{booktabs}
\usepackage{gensymb}
\usepackage{array}
\usepackage{caption}
\usepackage{multicol}
\usepackage{algpseudocode,algorithm,algorithmicx}
\usepackage{colortbl}

\newcommand{\rev}[1]
{{\color{blue}\textbf{#1}}}

\renewcommand{\theequation}{\arabic{equation}}
\usepackage{bm}

\graphicspath{ {./Fig/} }

\begin{document}
\title{AquaIntellect: A Semantic Self-learning Framework for Underwater Internet of Things Connectivity}
%
        

\author{{Ananya Hazarika and Mehdi Rahmati}\\
Department of Electrical and Computer Engineering, Cleveland State University, OH, USA  \\
Emails: 
{a.hazarika@vikes.csuohio.edu and m.rahmati@csuohio.edu}}
  

\maketitle

\begin{abstract}
The emerging paradigm of Non-Conventional Internet of Things~(NC IoT), which is focused on the usefulness of information as opposed to the notion of high volume data collection and transmission, will be an important and dominant part of human life in the near future.
This paper proposes a novel semantic-based approach for addressing the unique challenges posed by underwater NC IoT. We present an intelligent sensing strategy for exploring the semantics of the underwater environment by judiciously selecting the data to transmit, thereby minimizing redundancy for utmost relevant data transmission. We introduce an evolutionary function for the selection of the semantic-empowered messages relevant to the specific task within a minimum Age of Information~(AoI), a freshness metric of the collected information, and for monitoring the underwater environment for performance optimization. A DNN-empowered Bayesian integrated with an adaptive surrogate model optimization will determine the optimal placement strategy of the sensors and the uncertainty level of the underwater landscape. An Adaptive Expected Improvement (AEI) mechanism is introduced to predict the optimal arrival rate for enabling a synchronized data sensing and transmission ecosystem, ensuring efficiency and timeliness. Simulation results show that the proposed solution outperforms conventional approaches.
\end{abstract}

\IEEEpeerreviewmaketitle
\section{Introduction}
The imminent rollout of 5G and Beyond~(5GB) networks in the upcoming years is set to revolutionize the digital landscape, requiring integration of terrestrial and non-terrestrial networks, such as underwater IoTs, due to their ubiquitous nature.
Therefore, to fully exploit the potential of 5GB and ensure seamless and reliable coverage across varied terrains and vast geographical regions, a spectrum of complementary services and solutions is needed.  
The challenge is to standardize these technological advancements to ensure interoperability, scalability, and sustainability of different system components.
%
%
%
However, the lack of a standard for this heterogeneous ecosystem presents significant challenges and can act as a barrier to seamless interoperability, a cornerstone for any network to reach its full potential.
%
%
%
%
While terrestrial IoT has been the main focus of extensive research~\cite{al2015internet,xiao2014user}, the underwater NC IoT remains relatively uncharted. 
Here, nodes differ in terms of $(i)$ the type of sensed data, $(ii)$ the energy source for operation, and $(iii)$ the mode of data transmission and interaction among the nodes, considering energy and communication costs in this harsh environment~\cite{rahmati2019eco}. 
The challenge is further exacerbated when considering the unique requirements for underwater communications. 
While Radio Frequency~(RF) waves, optical waves, and magnetic induction have significant limitations underwater, acoustic waves seem to be the only possible solution for longer transmission ranges. Despite its ability to propagate over long distances, underwater acoustics come with their own set of challenges, including long transmission delay and variable sound speed due to changing water conditions with temperature, depth, and salinity.

Historically, performance in such networks was gauged using metrics like delay and successive time of delivery~\cite{bianchi2000performance}. However, these metrics often fall short in ensuring the freshness of transmitted information. Age of Information~(AoI) has emerged as a more holistic metric, quantifying the timeliness of received packets at the destination~\cite{yates2020age}.
Maintaining data freshness while ensuring interoperability among the nodes in NC IoTs is a pivotal challenge, especially in preventing resource under-utilization. Furthermore, the unpredictable nature of underwater environments necessitates the development of low-latency, robust configurations for efficient data acquisition and transmission. Bayesian optimization, known for its power in handling expensive-to-evaluate black-box functions, seems a promising avenue for addressing uncertainties in underwater sensing scenarios. To achieve intelligent sensing and transmission in underwater environments, it is essential to develop a model that deeply understands the complex dynamics in interactions, and efficiently and reliably extracts meaningful information under uncertainty.


In this paper, we propose a semantic-based self-learning approach  
that highlights the importance of observational context, asserting that the significance and context of a message often outweigh its simple occurrence, especially given the prevalent volatile transmission scenarios. We utilize a Bayesian optimization to navigate the uncertainties of underwater environments, ensuring efficient in-situ data acquisition and transmission. 
Our key contributions include:
\begin{itemize}
    \item envisioning a new underwater IoT design for underwater sensing, as traditional approaches are mostly not aware enough of the environment's behavior. This is achieved by interpreting the semantics of the updates, drawing from the concept of AoI, and employing a model predictive Deep Neural Network~(DNN)-based Bayesian network.
    \item enhancement of the process by focusing on the optimization of energy consumption across multiple subsets. This is achieved through a multi-task learning strategy that utilizes shared representations rooted in the semantics previously extracted.
\end{itemize}

\section{Related Works}\label{sec:rel}
%
Authors in~\cite{xiao2014user} elaborate on device interoperability in which large-scale cooperation between the IoT nodes and heterogeneous device subnets are required. The study presented in~\cite{fang2022average} demonstrates an active queue management approach in underwater wireless sensor networks to effectively lower the peak AoI and energy costs in underwater IoTs.
%
Syntactic interoperability in data being exchanged between two or more IoT system components with incompatible data formats or data structures is discussed in~\cite{noura2019interoperability}.
While semantic interoperability can appear to be similar to syntactic interoperability, semantic interoperability focuses on data being exchanged between and understood by two or more heterogeneous IoT ecosystem components~\cite{ganzha2018towards}. 
While authors in~\cite{popovski2020semantic} explore the possibilities of IoT focusing on minimizing communication overhead in Fog-Radio Access Networks (F-RANs) with semantics-aware grant-free radio access, the work in~\cite{lee2022query} tackles a critical challenge in uncertain data, specifically in adversarial attacks on discrete sequential data. The Blockwise Bayesian Attack~(BBA) framework presented in~\cite{lee2022query} introduces a method for generating adversarial examples, using a categorical kernel with ARD for dynamic position calculation and enhancing efficiency with block decomposition and history subsampling. 
\section{Proposed Solution}\label{sec:sol}

\begin{figure}[!t]
\includegraphics[width=3.45in]{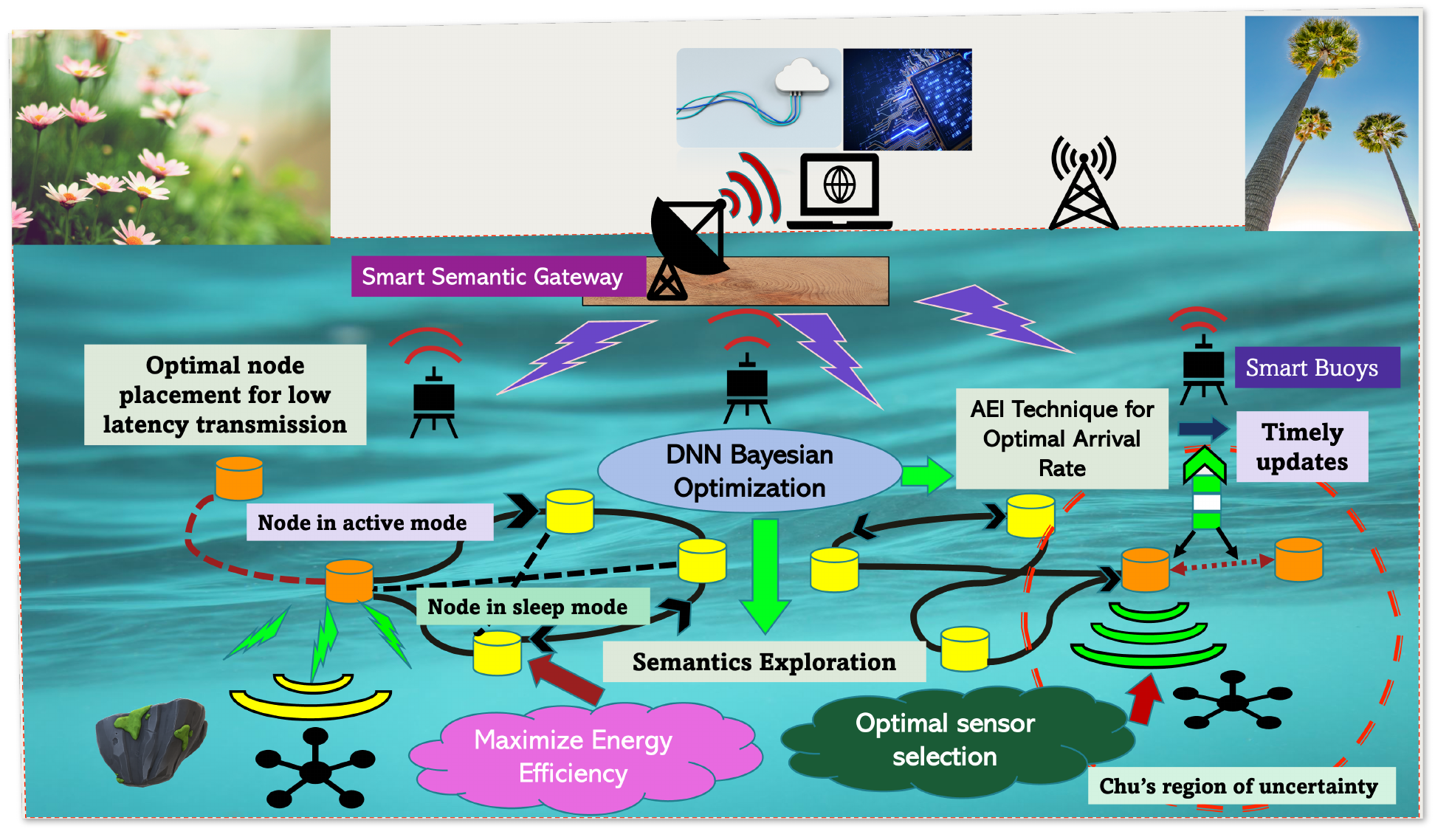}
\caption{The proposed terrestrial and underwater cohesive NC IoT, introducing smart buoys, called Smart Semantic Gateways~(SSGs). 
A semantics-driven approach interprets and prioritizes the information based on relevance and significance in the presence of uncertainties.}\label{fig:sysmodel}
\end{figure}

\textbf{System Model:} 
The proposed architecture, illustrated in Fig.~\ref{fig:sysmodel}, integrates terrestrial and underwater environments to create a cohesive NC IoT framework.
At the intersection of the two domains, Smart Semantic Gateways~(SSGs) oversee various heterogeneous underwater subnets, $S$ subnets in total, each comprising
$k_s$ nodes. These SSGs are adept at semantically processing the data as they intercept by fine-tuning terrestrial network parameters, enhancing the efficiency and speed of data collection and computation.
Diving into the underwater segment, the architecture includes underwater IoT nodes, designed to monitor the vast and unpredictable underwater environment without human intervention. These self-running nodes are equipped with underwater acoustic modems, ensuring robust communication even in the challenging underwater channel. The precision in data collection by these dynamic nodes is contingent on the localization of these nodes, which the architecture ensures. 
By processing data from underwater nodes in near-real-time, SSG buoys play a pivotal role in immediate optimization based on various factors such as node's working condition, data integrity, and transmission efficiency. Furthermore, these buoys aggregate data from multiple subnets, reducing redundant information and conserving bandwidth. Through the simultaneous integration of terrestrial and underwater segments, this novel architecture provides a comprehensive solution for monitoring and processing environmental data, ensuring that no land or water terrain is beyond the reach of modern IoT solutions.
\textbf{Semantic-driven AoI:} The underwater environment presents unique challenges for IoT nodes, especially in timely data acquisition and transmission regarding the significant propagation delays in underwater acoustic channels. 
We propose a semantics-driven approach for underwater IoT systems to interpret information semantics amidst uncertainties, enhancing data transmission and mitigating the overall delay by prioritizing information based on its relevance and significance. Here is a detailed overview of our proposed system framework for semantic extraction.

\subsubsection{Chu Spaces for Semantic Interpretation}
In underwater settings, sensor placements, often driven by engineering intuition rather than strategic design, which can result in inconsistent data and information gaps. 
To encapsulate and structure the uncertainty inherent in underwater environments, we adopt a pioneering model based on Chu spaces~\cite{pratt1995chu} for our NC IoT systems. 
This space is denoted over a set $\mathbf{C}$ by $H=(A,r,X)$ where $\{x_1,x_2,...,x_n\}$ are elements of $X$ that represent the IoT nodes involved in sensing useful information or messages for transmission on detecting a change in the environment. Let $A$ represent the set of all instances where the AoI surpasses a critical threshold $M$ such that $A_i$=$\Delta(t_i)>M$ for $i$ $\in$ $(1,2,...,n)$. Each element in $A_i$, i.e., $\{ A_1, A_2, \ldots, A_n \}$, corresponds to an AoI value for a specific status update sensed by the nodes in $X$ within their respective subnets. Essentially, if the AoI of a sensed status update from a node in $X$ exceeds $M$, it becomes part of the violation set $A$. In this defined Chu space, we do not impose any cardinality restrictions. This structured approach affords us a clearer understanding of the variability in message transmissions due to underwater environment dynamicity.
To delve deeper into the semantics and analyze the ambiguity intrinsic to the IoT system, we introduce the evolutionary function, $r:A\times X\rightarrow \mathbf{C}$, denoted by $r =\Pr(A|X)$ where $\Pr$ stands for the probability function. This function is instrumental in mapping the ever-evolving relationships between the AoI and the IoT nodes, thereby providing a robust framework for assessing and interpreting information in a dynamically shifting underwater environment. 
\subsubsection{Optimal Sensor Placement and Enhanced Energy Efficiency}  
The underwater environment introduces unique challenges, especially with the channels, which bring issues such as frequency-dependent path loss and significant propagation delays, just to name a few. To effectively tackle these challenges, the Underwater Acoustic~(UWA) channel model for deep sea can be empirically modelled by Urick attenuation formula~\cite{brekhovskikh1991fundamentals} $A_b(f,d)$ = $A_0 a(f)^d d^\zeta$.
Here \(A_b(f,d)\) denotes the absorption loss (in dB) with \(f\) representing frequency (in kHz), $A_0$ is the reference attenuation, $a(f)$ is Thorp's absorption coefficient, d is the distance between the transmitting and receiving units, and $\zeta$ is the cylindrical or spherical spreading factor. Note that to accurately predict attenuation in shallow water, the above formula should incorporate reflections, refraction from the water's bottom and surface, grazing angles, and bottom sediment density~\cite{rogers1981onboard}. 
The integration of the UWA channel model into our optimization framework provides a more realistic understanding of event detection in the underwater environment. 
Therefore, we propose to determine the optimal sensor count to improve system perceptibility by increasing the likelihood of event detection by IoT nodes within a specific subnet.
We articulate the probability of detecting a certain event at a Point of Interest (PoI) \(j\) by \(K\) sensors in \(X\), within a range \(l_{b_{k}}\) (where \(k\) spans from 1 to K) through a probabilistic model
\begin{equation}
    \Pr(X=k)= 
\begin{cases}
\frac{1}{k!}(\delta A_b(f,d_k))^{-k} e^{{-1}/({\delta A_b(f,d_k)})},& \text{} d_k\geq l_{b_k}\\
    1,& \text{} d_k<l_{b_k}.
\end{cases}
\end{equation}

Here $d_k$ is the distance in meters from the $k^{th}$ sensor and \(\delta\) is the decay factor. While the node in $X$ senses any environment change due to the occurrence of events in $X$ within the subnet ${b_k}$ which has a radius of $l_{b_k}$, it immediately sets the probability to $1$ and transmits the status updates of $X$. However, in densely deployed networks, only a subset of sensors can achieve successful detection based on their geographical conditions. Our proposal aims to determine the optimal number of sensors that can detect events effectively while also considering an efficient wake-up probability to better understand the semantics of underwater phenomena. Using the optimal number of sensors, the expectation of $X$ to obtain a successful detection by setting the wake-up probability of each of the sensors to $\gamma_{wake}$ is denoted by
 \begin{equation}
E(X) = \sum_{k\in K} \Bigg( \big(1-(1-\gamma_{wake} \epsilon_k)^k\big) \Pr(X=k) \Bigg).
\label{wakeupscheme}
\end{equation}

Here, \(\epsilon_k\) determines the efficiency of the \(k^{th}\) sensor in detecting events.
We define an optimization problem, denoted as \textbf{P1} to determine the optimal number of $K$ sensors that should be active in a given cycle to maximize the probability of sensing valuable information in a cost-effective manner. Mathematically
\begin{align} 
\textbf{P1}: \max_{K}\quad & E(X),\\
\text{s.t. }\ \quad& \gamma_{wake}\leq \gamma. \nonumber 
\end{align}

Solving \textbf{P1} provides the optimal number and spacing of sensors denoted by $K^{\ast}$ and $d_k^{\ast}$ respectively for thorough event detection. Once an event is identified, it is essential for these sensors to stay active, ensuring timely data relay to surface nodes without surpassing the AoI threshold.
We store the optimal location of $K$ sensors responsible for successful detection in $X$ Chu spaces.
\subsubsection{Quantifying Freshness for Active Transmission of Sensed Updates}
In underwater IoT landscape, propagation delays pose a prominent challenge. As nodes continuously sense changes, it is crucial they transmit the most recent data to the surface buoy. 
Since the high uncertainty or fluctuation of underwater information makes the average AoI unreliable, a statistical guarantee can be ensured in such a way that the AoI will not exceed a threshold for a certain percentage of the time. When the AoI exceeds a certain threshold, it is known as the violation probability of AoI \cite{franco2020extended}. The violation probability of AoI should be minimized to ensure that the information sent by the IoT nodes is deemed to be too old. In underwater IoT environments, a First-Come-First-Served (FCFS) queuing model is preferred over Last-In-First-Out (LIFO) due to its ability to provide more predictable and stable data flow amid the high uncertainty and fluctuation of underwater conditions~\cite{llor2012underwater}. FCFS, combined with AoI analysis, ensures that all data packets are processed in a timely manner, balancing data freshness with fairness, and offering a statistical guarantee against excessive information aging, crucial in such challenging environments. We define $r$ as the function of AoI in FCFS \textbf{M/M/1} queues along with the sensing information of the event $X$~\cite{9324753} denoted by 
\begin{equation}
    r= \Pr(A|X)= \pi_{s} \Pr(X=k),
\end{equation}
where $A_i$ is the violation probability in $A$ as \textbf{M/M/1} queuing systems~\cite{9324753} denoted by
\begin{equation}
\begin{split}
 A_i=Pr\{\Delta(t_i)>M\}=&e^{-(\mu-\lambda)M} + (\frac{\mu}{\lambda-\mu}-\lambda M)e^{-\mu M}\\
 & -\frac{\mu}{\lambda-\mu}e^{-\lambda M}.
 \label{eq3}
 \end{split}
 \end{equation} 

 In~\eqref{eq3}, while $\lambda$ signifies the arrival rate of status updates, its precise behavior within the system remains ill-defined. The service rate is denoted by $\mu$ and $\pi_{s}$=$\lambda A_i e^{-\lambda A_i}$ is the probability that the status contains the information or updates of the event X for $i$$\in$$X$ until the limit set by violation age of probability $A$~\cite{KUMAR2008187}.  
As \(r\) lacks gradient data and clarity on the arrival rate \(\lambda\), we categorize it as a black-box objective function. 

\textbf{DNN-Based Bayesian Optimization for Sensor Selection and Transmission:}
There arises multiple challenges in navigating the underwater nodes when the task at hand demands simultaneous optimization of both the quantity and spatial distribution of sensors. To adeptly maneuver through this multi-dimensional challenge, we have fine-tuned our semantics-aware Bayesian optimization to accommodate the spatial variation associated with sensor distribution. The main goal of this optimization is to find the best configuration \(X^*\) that optimally amplifies \(E(X)\), a configuration that seamlessly integrates both the sensor count and their precise geolocations. Our approach is framed mathematically as follows.
\begin{equation}
  X^{\ast} = \text{argmax}\ E(X),
\end{equation}
where $X^{\ast}$ not only provides information on the number of sensors but also their strategic positioning within the uncertainty domain.
We employ a Multiple Transmission Enabled Bayesian Optimization (MTBO) strategy to deduce the optimal arrival rate of the black-box objective function $r$. This rate ensures the freshness of information in the underwater IoT network while concurrently facilitating data transmission from the selected sensors. MTBO's objective is to discover the function $r$'s minimizer
\begin{equation}
  \lambda^{\ast} = \text{argmin} (r(\lambda)). 
\end{equation}

The Bayesian optimization solution~\cite{frazier2018bayesian} adopts a dual-fold strategy based on the surrogate model, tailored to tackle sensor optimization challenges by amplifying the sampling potency and capturing the true essence of our objective, especially when the geographical coordinates of sensors are considered.  Subsequently, we hone the acquisition function, leveraging the surrogate model's insights for sensor selection and transmission efficiency. 
\subsubsection*{1) Bayesian Surrogate Modeling with Gaussian Process and DNN Integration}
In order to comprehend the intricate dynamics of the arrival rates in Chu's vast and unpredictable space, we rely on adaptive surrogate modeling and Bayesian optimization techniques~\cite{lei2021bayesian,frazier2018bayesian}. To better navigate and gauge confidence in black-box objective functions, we transition these functions into a probabilistic framework. Specifically, we employ Gaussian processes to capture the distribution of samples, thereby offering insights into their covariance and mean. The selection of the Gaussian process as our foundation arises from its computational efficiency and exceptional capability in addressing uncertainty. Given the computational constraints of underwater nodes and the vastness of the search domain, the Gaussian process acts as a resilient backbone. We strategically model the surrogate using a DNN, which is expertly crafted to map Chu spaces. With a DNN-based surrogate model in place~\cite{9841555}, the challenges posed by randomized input parameter selections in simulations are effectively managed.
Training our surrogate model necessitates the partitioning of data into distinct training and testing subsets. These datasets facilitate predictions related to the optimal number and positioning of sensors. Once the most effective sensor locations within $X$ are identified, our methodology evolves. It uses this newly acquired information to fine-tune the evolutionary function $r$, maximizing the efficiency of multiple transmissions from $X$ and minimizing latency. To support our methodology, we make an assumption that the distribution for the arrival rate follows a Gaussian pattern. This Gaussian distribution's mean and variance are expressed as $\lambda$=$\textbf{N}(\mu_{a},\Sigma)$.
We further emphasize the proficiency of Gaussian Process (GP) based surrogate models, noting their distinction as some of the most computationally efficient tools in simulation-based design optimization and uncertainty analysis. We formulate the standard GP for our evolutionary function \(r\), expressed as
\begin{equation}
    Y_{G}(\lambda)=r(\lambda)+s(\lambda),
\end{equation}
where \(s(\lambda)\) symbolizes the Gaussian stochastic process noise, defined by a zero mean and variance \(\sigma^{2}\).
For the formulation of the standard GP, we operate under the assumption that all variables are fully observed. 
An integral part of our methodology is the adaptive sampling approach to monitor the evolving minimum of the dynamic BO function, ensuring efficient surrogate model development. By iteratively selecting new training sample points and refining the surrogate model, this approach hones the model until the desired accuracy level is achieved. Hence, this iterative strategy meticulously tracks the optimal \(\lambda^{\ast}\) to guarantee minimal delay at the precise time instant \(t^{\ast}\). 
\begin{figure*}[!t]
\centering
\includegraphics[width=2.25in]{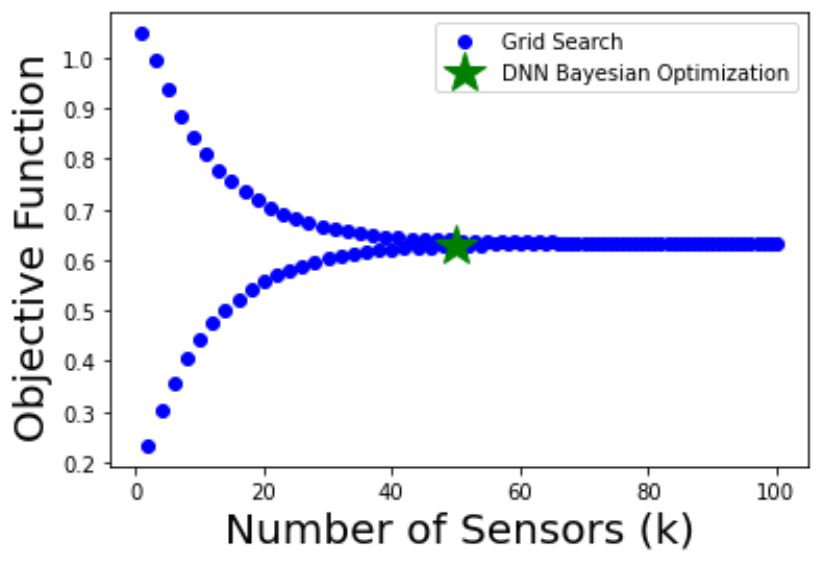} \hspace{-2mm}
\includegraphics[width=2.25in]{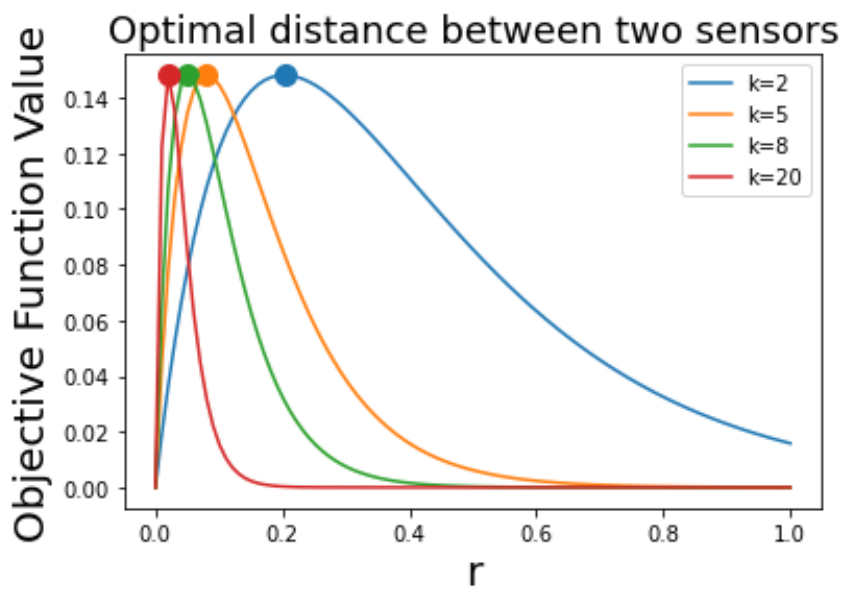} \hspace{-2mm}
\includegraphics[width=2.57in]{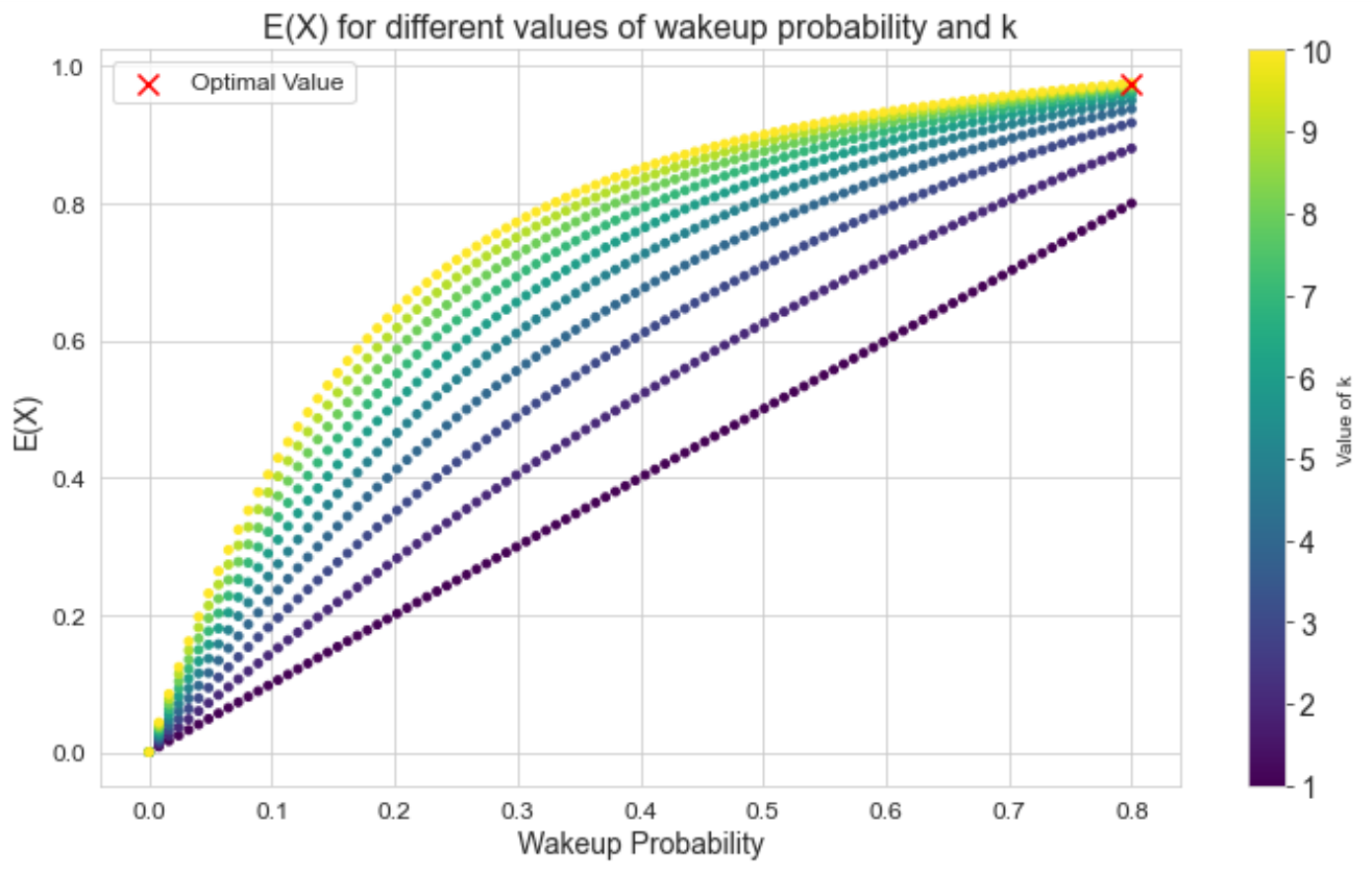}
\caption{(left) The optimal number of sensors from DNN-based Bayesian Optimization for a search space of huge uncertainty; (center) The distance between two sensors for an optimal sensor placement with maximum sensing information; (right) Visualization of $E(X)$ values across varying wakeup probabilities for a set of sensors.}
\vspace{-0.1in}
\label{fig:optimalplacement}
\end{figure*}
\subsubsection*{2) Adaptive Expected Improvement for Enhanced Sampling Prediction}
Based on the foundational knowledge of BO, the significance of the acquisition function, $a(\lambda)$, is indispensable for robust sampling in an underwater architecture~\cite{wilson2018maximizing}. Derived from the posterior distribution information of the GP at a particular input locus, it orchestrates the search for the optimal arrival rate $\lambda^{\ast}$ for each subnet. Traditionally, this function is driven by the Expected Improvement (EI), aiming to maximize the improvement over a predetermined threshold, \(c\), which signifies the best-observed value of the objective function \(r\). This is shown by a utility function given by $I(\lambda)$=$(r(\lambda)-c)\textbf{I}(r(\lambda)< c)$. The term r\( ((\lambda) - c) \) calculates the improvement of the r(\( \lambda \)) over the current best value \( c \) and \( \mathbf{I}(r(\lambda) > c) \) is an indicator function that equals 1 if \( r(\lambda) \) is greater than \( c \), and 0 otherwise. This ensures that the utility function yields a positive value only when a new sample \( \lambda \) leads to an objective function value \( r(\lambda) \) that is better than the current best \( c \). Given the inherent volatility and complexity of underwater environments, the Adaptive Expected Improvement (AEI) approach addresses these underwater complexities. Distinct from the rigid threshold typical of conventional EI, AEI offers flexibility by dynamically adapting its threshold, taking into account real-time sensor feedback, and continuously refining its understanding of Chu spaces by $c_t$ $=$ $c$ + $\phi(\lambda_t)$
where $\lambda_t$ represents the incoming sensor data at time $t$ and  $\phi$ quantifies the change in threshold. AEI is then formulated as
\begin{equation}
    AEI(\lambda) = (r(\lambda) - c_t) \textbf{I}(r(\lambda) < c_t).
\end{equation}

This dynamic recalibration ensures that our optimization is not anchored to outdated thresholds, leading to a more responsive search for the optimal arrival rate, \(\lambda^{\ast}\). 
 We have incorporated a feedback loop for continual refinement, allowing for the recalibration of the threshold \(c\) whenever there is a significant discrepancy, denoted as $\delta$, between the model's predicted optimal rate and the observed actual outcomes. The discrepancy at time $t$ is calculated as
\begin{equation}
    \delta_t = |predicted(\lambda_t) - actual(\lambda_t)|.
\end{equation}

Subsequently, the threshold for the next iteration is adjusted based on this discrepancy using the recalibration factor \(\omega\) as follows.
\begin{equation}
    c_{t+1} = c_t + \omega \cdot \delta_t.
\end{equation}
This approach ensures that the threshold \(c\) is dynamically updated to reflect the latest data, thereby enhancing the model's accuracy and responsiveness.
\begin{figure}[!t]
\centering
\label{fig:delay}
\includegraphics[width=3.4in]{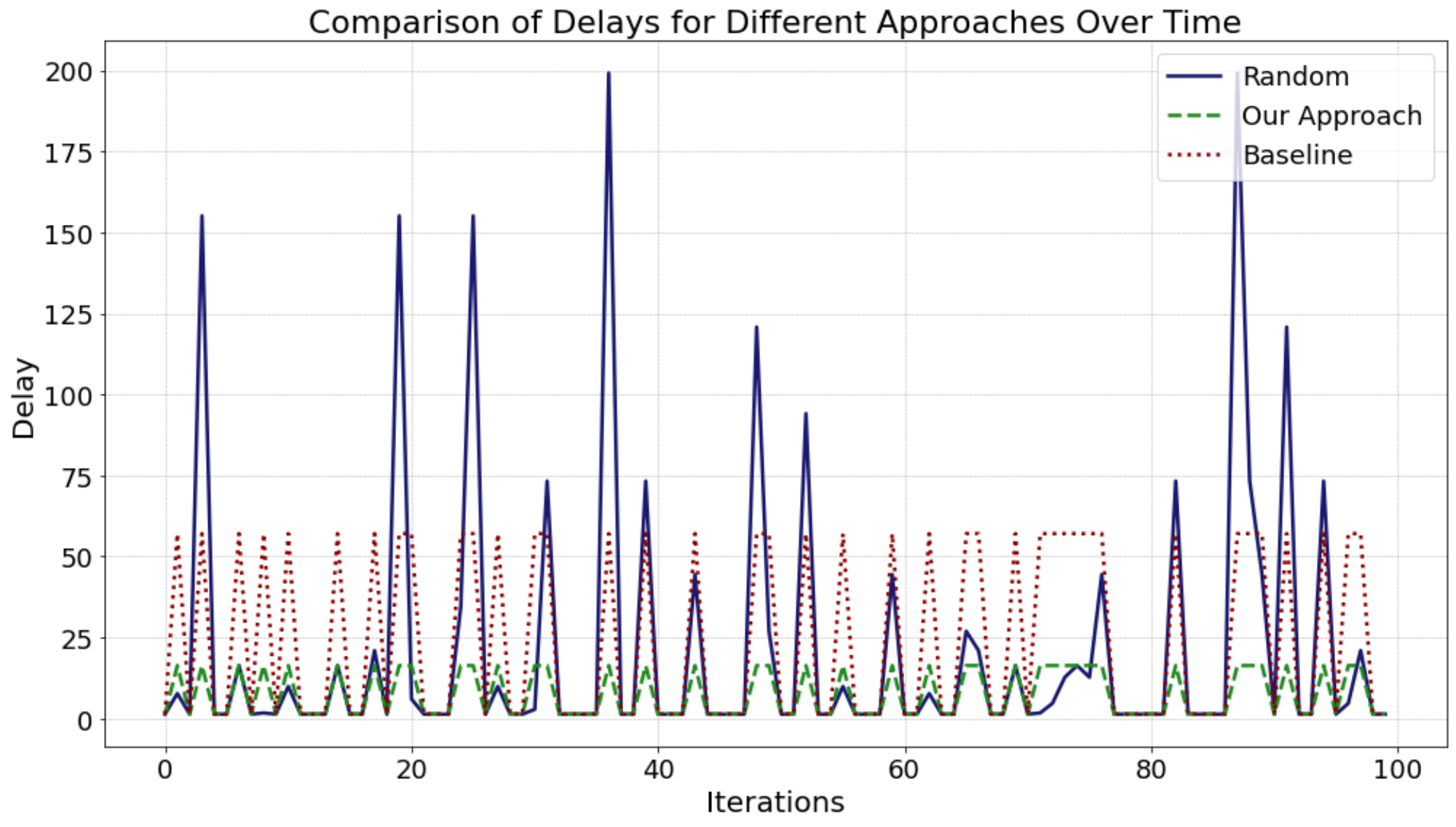}
\caption{Comparative analysis of delays over time for our proposed approach with random and non-adaptive placement strategies.}
\label{fig:delay}
\end{figure}
\begin{figure*}[!t]
 \centering
\includegraphics[width=3.8in]{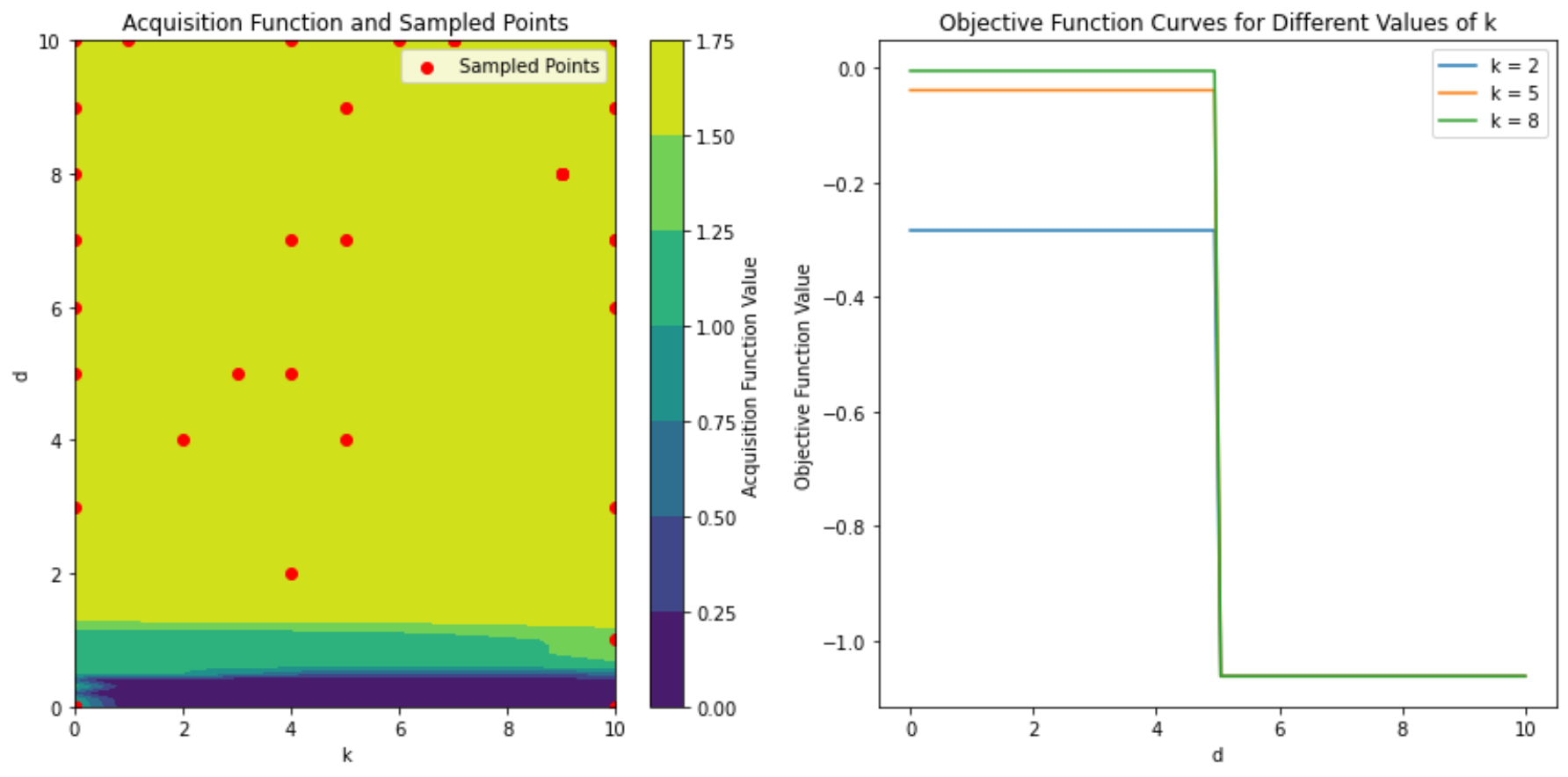}
\includegraphics[width=3.2in]{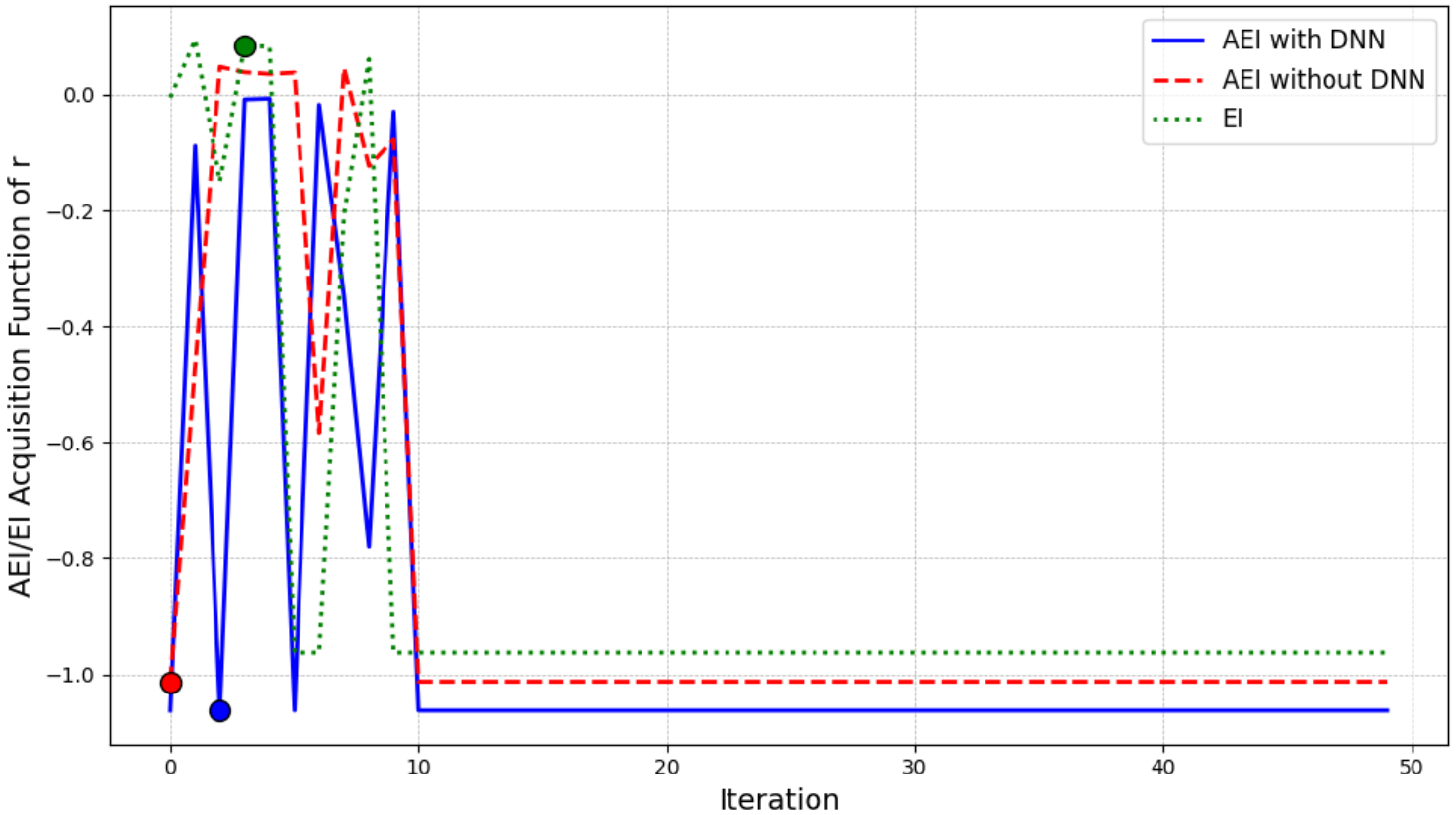}
\caption{(left) The acquisition function from Bayesian Optimization~(BO) for the optimal number of sensor selections with their respective distance $d$; (center) The behavior of the surrogate modeling-based  BO of $r$ to determine the optimal values of $k$; (right) The comparative efficiency of the AEI method over the EI approach in achieving optimal arrival rates for underwater IoT subnets.}
\label{fig:acquisitionfunc1}
\end{figure*}
\section{Performance Evaluation}\label{sec:eval}
\textbf{Simulation Setup:}
For our simulations, we initialized \(\mu\) at 1, \(\lambda\) at 0.8, \(K\) at 2, and the probability of system wake-up \(\gamma_{wake}\) at 0.9, with decay parameter \(\delta\) at 0.6. The reference boundary distance \(l_{b_{k}}\) was 5 m. We generated an initial sample of 10 points, expanding to 100 samples per iteration for optimization. Our DNN architecture consisted of an input and two hidden layers with 64 neurons each, employing ReLU activation and dropout of 0.5 to prevent overfitting. The Adam optimizer with a learning rate of 0.01 minimized the mean squared error loss. BO was performed over 40 iterations in three scenarios: with a DNN, without a DNN, and using EI. To visually distinguish the methodologies, we adjusted the AEI without DNN and EI results upwards by 0.05 and 0.1, respectively.

\textbf{Simulation Results:} Fig.~\ref{fig:optimalplacement}(left) illustrates the optimal number of energy-efficient sensors in wakeup mode, while Fig.~\ref{fig:optimalplacement}(center) shows the optimal distances for a fixed set of $k$ sensors from DNN BO. In Fig.~\ref{fig:optimalplacement}(center), DNN BO converges more rapidly at 48 sensors with higher intrusion detection performance than Grid Search at 70 sensors, resulting in energy savings and reduced data redundancy. Figure~\ref{fig:optimalplacement}(right) visualizes the impact of varying wakeup probabilities and sensor counts (\(k\)) on the expected value \(E(X)\) in the underwater IoT context. The distinct red curve in Fig.~\ref{fig:optimalplacement}(right) highlights the peak efficiency, achieved at a wakeup probability of 0.8 with 10 sensors, guiding towards the most efficient configuration for the underwater IoT system. Figure~\ref{fig:delay} illustrates a comparison between our optimal sensor placement approach and the traditional random and baseline strategies, where the baseline employs a fixed number of sensors, highlighting the efficacy of our method in reducing delays. The graph in Figure~\ref{fig:delay} demonstrates the superiority of our DNN BO approach with its consistently lower delay times, effectively illustrating the efficiency and reliability of our placement method. Figure~\ref{fig:acquisitionfunc1} (left) showcases the acquisition function, which is guided by insights from a surrogate model. This function efficiently guides the selection of sensors and transmission rates within the parameter space ($K^{\ast}$, $d_k^{\ast}$). The colored contour plot in Fig.~\ref{fig:acquisitionfunc1}(left) visually represents the acquisition function, with higher contour levels indicating regions of increased uncertainty and potential improvement. Figure~\ref{fig:acquisitionfunc1}(left) illustrates the optimization process's ability to explore and exploit the parameter space effectively for sensor placement and transmission rate determination. Figure~\ref{fig:acquisitionfunc1}(center) shows the behavior of r across different spatial configurations for selected values of $k$ and a sensor configuration with \( K^{\ast} = 8 \) is seen to have the highest objective function value across the evaluated distances, indicating its potential optimality in the given figure. 
 Figure~\ref{fig:acquisitionfunc1}(right) shows the effective sampling prediction by achieving the optimized arrival rate of each underwater IoT subnet using our proposed AEI approach and compare with EI approach for effective transmission of information with lower latency. It can be clearly seen from Fig~\ref{fig:acquisitionfunc1}(right) that AEI approach converges faster in achieving the optimal arrival rate compared to the EI approach.

\section{Conclusion and Future Work} \label{sec:conc}
We introduced a semantics-aware self-learning framework tailored for exploring uncertainty in data acquisition in underwater environments. We employed an intelligent sensing strategy utilizing Chu spaces for the selection of the optimal sensor nodes with their specific destination for selective transmission of pivotal data with lower latency. To further enhance our placement strategy, Bayesian Optimization was incorporated, leveraging an adaptive Gaussian surrogate model. This model efficiently sampled the underwater environment, predicting uncertainty levels and significantly minimizing data redundancy. Drawing on the principles of surrogate sampling, we proposed the AEI method for determining the most suitable data arrival rate, thus ensuring timely sensing and data relay. The integration of these strategies resulted in a seamless sensing-transmission workflow, establishing an environment suitable to prompt and significant data communication. In future work, we plan to evaluate the proposal in a real experimental setting, varying water depth and environmental conditions.

\textbf{Acknowledgements:} This work was supported by the NSF Award No. 2215388.
\balance
\bibliographystyle{IEEEtran}
\bibliography{VCC_Final}
\end{document}